\documentclass[journal]{IEEEtran}

\ifCLASSINFOpdf
\else
\fi

\usepackage{setspace}
\usepackage{graphicx,graphics,epsfig,float}
\usepackage{amsmath} 
\usepackage{amssymb}  
\usepackage{mathrsfs}
\usepackage{enumerate}
\usepackage{subfigure}
\usepackage{colortbl}
\usepackage{bm}
\usepackage{psfrag}
\usepackage{cite}
\usepackage{color}

\usepackage{bm}

\DeclareMathAlphabet{\mathsfbr}{OT1}{cmss}{m}{n}
\SetMathAlphabet{\mathsfbr}{bold}{OT1}{cmss}{bx}{n}
\DeclareRobustCommand{\msf}[1]{%
  \ifcat\noexpand#1\relax\msfgreek{#1}\else\mathsfbr{#1}\fi
}

\makeatletter
\newcommand{\msfgreek}[1]{\csname s\expandafter\@gobble\string#1\endcsname}
\makeatother

\DeclareFontEncoding{LGR}{}{} 
\DeclareSymbolFont{sfgreek}{LGR}{cmss}{m}{n}
\SetSymbolFont{sfgreek}{bold}{LGR}{cmss}{bx}{n}
\DeclareMathSymbol{\salpha}{\mathord}{sfgreek}{`a}
\DeclareMathSymbol{\sbeta}{\mathord}{sfgreek}{`b}
\DeclareMathSymbol{\sgamma}{\mathord}{sfgreek}{`g}
\DeclareMathSymbol{\sdelta}{\mathord}{sfgreek}{`d}
\DeclareMathSymbol{\sepsilon}{\mathord}{sfgreek}{`e}
\DeclareMathSymbol{\szeta}{\mathord}{sfgreek}{`z}
\DeclareMathSymbol{\seta}{\mathord}{sfgreek}{`h}
\DeclareMathSymbol{\stheta}{\mathord}{sfgreek}{`j}
\DeclareMathSymbol{\siota}{\mathord}{sfgreek}{`i}
\DeclareMathSymbol{\skappa}{\mathord}{sfgreek}{`k}
\DeclareMathSymbol{\slambda}{\mathord}{sfgreek}{`l}
\DeclareMathSymbol{\smu}{\mathord}{sfgreek}{`m}
\DeclareMathSymbol{\snu}{\mathord}{sfgreek}{`n}
\DeclareMathSymbol{\sxi}{\mathord}{sfgreek}{`x}
\DeclareMathSymbol{\somicron}{\mathord}{sfgreek}{`o}
\DeclareMathSymbol{\spi}{\mathord}{sfgreek}{`p}
\DeclareMathSymbol{\srho}{\mathord}{sfgreek}{`r}
\DeclareMathSymbol{\ssigma}{\mathord}{sfgreek}{`s}
\DeclareMathSymbol{\stau}{\mathord}{sfgreek}{`t}
\DeclareMathSymbol{\supsilon}{\mathord}{sfgreek}{`u}
\DeclareMathSymbol{\sphi}{\mathord}{sfgreek}{`f}
\DeclareMathSymbol{\schi}{\mathord}{sfgreek}{`q}
\DeclareMathSymbol{\spsi}{\mathord}{sfgreek}{`y}
\DeclareMathSymbol{\somega}{\mathord}{sfgreek}{`w}

\DeclareMathSymbol{\svarsigma}{\mathord}{sfgreek}{`c}

\DeclareMathSymbol{\sGamma}{\mathalpha}{sfgreek}{`G}
\DeclareMathSymbol{\sDelta}{\mathalpha}{sfgreek}{`D}
\DeclareMathSymbol{\sTheta}{\mathalpha}{sfgreek}{`J}
\DeclareMathSymbol{\sLambda}{\mathalpha}{sfgreek}{`L}
\DeclareMathSymbol{\sXi}{\mathalpha}{sfgreek}{`X}
\DeclareMathSymbol{\sPi}{\mathalpha}{sfgreek}{`P}
\DeclareMathSymbol{\sSigma}{\mathalpha}{sfgreek}{`S}
\DeclareMathSymbol{\sUpsilon}{\mathalpha}{sfgreek}{`U}
\DeclareMathSymbol{\sPhi}{\mathalpha}{sfgreek}{`F}
\DeclareMathSymbol{\sPsi}{\mathalpha}{sfgreek}{`Y}
\DeclareMathSymbol{\sOmega}{\mathalpha}{sfgreek}{`W}

\DeclareRobustCommand{\mcal}[1]{%
  \ifcat\noexpand#1\relax\mathnormal{#1}\else\cal{#1}\fi
}
\DeclareRobustCommand{\BM}[1]{%
  \ifcat\noexpand#1\relax\bm{\boldUppercaseItalicGreek{#1}}\else\bm{#1}\fi
}
\makeatletter
\newcommand{\boldUppercaseItalicGreek}[1]{\csname var\expandafter\@gobble\string#1\endcsname}
\makeatother
\newcommand{\rv}[1]{\MakeLowercase{\msf{#1}}}  
\newcommand{\RV}[1]{\bm{\MakeLowercase{\msf{#1}}}}

\newcommand{\RS}[1]{\MakeUppercase{\msf{#1}}}

\newcommand{\Set}[1]{\mcal{#1}}

\DeclareMathOperator*{\argmin}{argmin}

\DeclareMathOperator*{\worst}{worst}

\DeclareMathOperator{\tr}{tr}
\DeclareMathOperator{\Leb}{\Lambda}

\newcommand{\anchor}{\ensuremath{\RS{A}}}
\newcommand{\pos}{\RV p}
\newcommand{\bpos}{\RV p^*}
\newcommand{\hpos}{\hat{\RV p}}
\newcommand{\UR}{\ensuremath{\mathscr{A}_{\mathrm{U}}}}
\newcommand{\dis}{\mathrm{dis}}
\newcommand{\syn}{\mathrm{clk}}
\newcommand{\noi}{\rv{n}}

\newcommand{\pmax}{\ensuremath{\pos_{\mathrm{max}}}}
\newcommand{\pmin}{\ensuremath{\pos_{\mathrm{min}}}}
\newcommand{\thr}{\ensuremath{\mathrm{th}}}
\newcommand{\Je}{\ensuremath{\bf{J}_{\text{e}}}}	
\newcommand{\lambdan}{\ensuremath{{\lambda}_n}}
\newcommand{\lambdaz}{\ensuremath{{\lambda}_0}}
\newcommand{\thetans}{\ensuremath{\rv{\theta}_n}}

\newcommand{\thetaks}{\ensuremath{\rv{\theta}_k}}
\newcommand{\thetaijs}{\ensuremath{\rv{\theta}_{ij}}}
\newcommand{\thetaizjzs}{\ensuremath{\rv{\theta}_{i_0j_0}}}

\newcommand{\thetak}{\ensuremath{\rv{\theta}_k^*}}
\newcommand{\thetaij}{\ensuremath{\rv{\theta}_{ij}^*}}
\newcommand{\thetaizjz}{\ensuremath{\rv{\theta}_{i_0j_0}^*}}

\newcommand{\thetanp}{\ensuremath{\rv{\theta}_n'}}
\newcommand{\thetaip}{\ensuremath{\rv{\theta}_i'}}
\newcommand{\thetajp}{\ensuremath{\rv{\theta}_j'}}
\newcommand{\thetakp}{\ensuremath{\rv{\theta}_k'}}
\newcommand{\thetaijp}{\ensuremath{\rv{\theta}_{ij}'}}
\newcommand{\thetaizjzp}{\ensuremath{\rv{\theta}_{i_0j_0}'}}

\newcommand{\sigdis}{\ensuremath{\sigma_{\text{dis}}}}
\newcommand{\sigsyn}{\ensuremath{\sigma_{\text{clk}}}}
\newcommand{\SPEB}{\ensuremath{\mathcal{P}(\pos)}}
\newcommand{\SPEBbest}{\ensuremath{\mathcal{P}_0}}
\newcommand{\Outage}{\ensuremath{\mathcal{O}}}
\newcommand{\POutage}{\ensuremath{\mathcal{O}_{\mathrm{aux}}}}
\newcommand{\PCR}{\ensuremath{\mathscr{A}_{\mathrm{AC}}}}
\newcommand{\CR}{\ensuremath{\mathscr{A}_{\mathrm{C}}}}

\newcommand{\Ia}{\ensuremath{\mathcal{I}_1}}
\newcommand{\Ib}{\ensuremath{\mathcal{I}_2}}
\newcommand{\Ic}{\ensuremath{\mathcal{I}_3}}

\newcommand{\Xcover}{\ensuremath{\rv{X}}}
\newcommand{\Sset}{\ensuremath{\rv{S}}}
\newcommand{\Y}{\ensuremath{\rv{Y}}}
\newcommand{\K}{\ensuremath{\rv{K}}} 
\newcommand{\X}{\ensuremath{\RV{X}}} 
\newcommand{\omg}{\ensuremath{\bm{\omega}}} 

\newcommand{\Nset}{\Set{N}}
\newcommand{\Ith}{\ensuremath{\mathscr{D}_\delta}}

\newcommand{\df}{\ensuremath{\thinspace\mathrm{d}}}

\newtheorem{theorem}{Theorem}
\newtheorem{proposition}{Proposition}
\newtheorem{definition}{Definition}

\begin{document}
\title{On the Outage Probability of Localization in Randomly Deployed Wireless Networks}
\author{Fengyu Zhou,~\IEEEmembership{Student Member,~IEEE}, and Yuan Shen,~\IEEEmembership{Member,~IEEE}
\thanks{This research was supported, in part, by the National Natural Science Foundation of China under Grant 61501279 and 91638204.

F. Zhou was with the Department of Electronic Engineering, Tsinghua University,
Beijing 100084, China. He is now with the Department of Electrical Engineering,
California Institute of Technology, Pasadena, CA 91125 USA (e-mail: f.zhou@caltech.edu).

Y. Shen is with the Department of Electronic Engineering, and Tsinghua National Laboratory for Information Science and Technology, Tsinghua University, Beijing 100084, China (e-mail: shenyuan\_ee@tsinghua.edu.cn).
}
}
\maketitle

\begin{abstract}
This paper analyzes the localization outage probability (LOP), the probability that the position error exceeds a given threshold, in randomly deployed wireless networks. Two typical cases are considered: a mobile agent uses all the neighboring anchors or select the best pair of anchors for self-localization. We derive the exact LOP for the former case and tight bounds for the LOP for the latter case.
The comparison between the two cases reveals the advantage of anchor selection in terms of LOP versus complexity tradeoff, providing insights into the design of efficient localization systems.

\end{abstract}


%
\IEEEpeerreviewmaketitle


\section{\label{Sec1}Introduction}

\IEEEPARstart{H}{igh}-accuracy localization is gaining its popularity in various applications nowadays, but traditional localization techniques exhibit their limitations in certain aspects \cite{WinConMazSheGifDarChi:J11}. For instance, the global navigation satellite system (GNSS), the most widely-used localization methods, often cannot meet the accuracy requirements in hash environments (e.g., indoors or underground). To this end, wireless local networks are introduced to complement existing techniques for enhancing localization performance.

Wireless localization networks consist of anchors and agents. Anchors have precisely known positions, while the agents measure the distances to neighboring anchors by wireless transmission and then infer their positions by solving a set of (overdetermined) equations formed by the inter-node distances and anchors' positions \cite{WinConMazSheGifDarChi:J11}. In this case, there exists a tradeoff for making distance measurements to more anchors, as it consumes additional power and time in return for diminishing improvement in localization performance.

Besides the localization accuracy, the localization outage probability (LOP) is also an important performance metric, which evaluates the probability that the position error exceeds a given threshold  \cite{Maz:04}.
To the best of the authors' knowledge, most related studies focused on the LOP of a mobile agent with known anchor deployment, and no analytical characterizations of the LOP have been presented.
Inspired by stochastic geometry approaches in wireless communications and localization \cite{WinPinShe:J09,BacBla:09a}, we adopt an alternative view to evaluate the LOP of an agent in randomly deployed wireless networks.


This paper considers the cases that all anchors participate in the localization and that the best pair of neighboring anchors is selected, and evaluates the localization performance in term of LOP. Comparison between these two cases demonstrates the benefit of anchor selection for efficient wireless localization.

\begin{figure}
\centering
\includegraphics{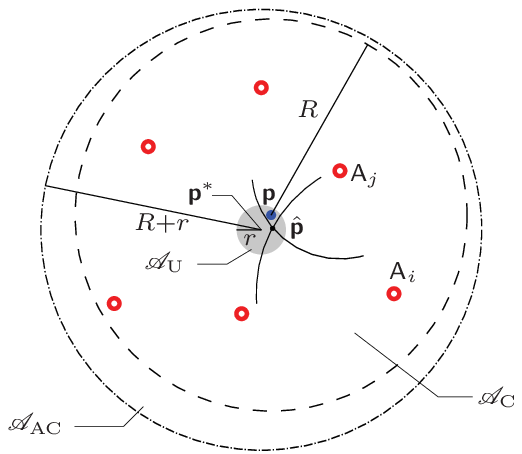}
\caption{System model: The grey circular $\UR$ (centered at $\bpos$) with radius $r$ denotes the uncertainty region of the agent. The true and estimated positions of the agent are denoted by $\pos$ and $\hpos$. The dashed circle $\CR$ (centered at $\pos$) denotes the communication region of the agent, while the dot dash circle $\PCR$ (centered at $\bpos$) denotes the auxiliary communication region.}
\label{figc1}
\end{figure}

\section{\label{Sec2}Problem Formulation}
\subsection{\label{Sec2Sub1}System Model}

Consider an agent with prior knowledge of its position $\pos$ locating in a small uncertainty region (UR) $\UR$, i.e., $\pos\in\UR$ (see Fig.~\ref{figc1}).\footnote{In most applications such as navigation, $\UR$ can be derived from the prior position and the maximum displacement of the agent within a time interval. Such a region is a rough estimation of agent's position {\it before} wireless localization, and it is usually much larger than the desired position error.}
For simplicity, let $\UR:=\{{\bf x}|{\bf x}\in\mathbb{R}^2, \Vert {\bf x}-{\bpos} \Vert \leqslant r\}$, where ${\bpos}$ is the center of the UR.
Moreover, we consider a circular region $\CR:=\{{\bf x}|{\bf x}\in\mathbb{R}^2, \Vert {\bf x}-{\pos} \Vert \leqslant R\}$ as the communication region of the agent, where $R$ is the maximum ranging distance of the agent, and there are $N$ anchors ${\anchor}_1, {\anchor}_2, ..., {\anchor}_N$ (with known positions)
uniformly and independently distributed in $\CR$. The distance and angle from anchor $k$ to $\bpos$ are denoted by $\rv{r}_k$ and $\thetak$, respectively, for $k\in\Set{N}$$:=\{1, 2, ..., N\}$.
The agent sends ranging requests to a specific subset of $\Nset$ and measures the round-trip time based on their replies for position estimation.



%
We consider line-of-sight (LOS) scenarios, where the range measurement ${\hat{\rv{r}}_\dis}$ can be modeled as
\begin{align}\label{ranging_exp}
{\hat{\rv{r}}_\dis}=r_\dis+\noi_\dis+\noi_\syn
\end{align}
where ${r_{\dis}}$ is the true distances between the agent and the anchor, and the noises $\noi_{\dis}$ and $\noi_{\syn}$ are modeled as two independent zero-mean Gaussian random variables with variances
$\sigma_{\dis}^2$ and $\sigma_{\syn}^2$, respectively. The former accounts for the ranging error from the time-of-arrival measurements, which depends on the communication distance $r_{\dis}$, while the latter accounts for the error from the clock drifts of the agent and the anchor.



For later derivations, we introduce the angle from anchor ${\anchor}_k$ to the agent (located at $\pos$) as $\thetaks$. The included angles of ${\anchor}_i$-$\pos$-${\anchor}_j$ and ${\anchor}_i$-$\bpos$-${\anchor}_j$ are denoted by $\thetaijs$ and $\thetaij$, respectively.

\subsection{\label{Sec2Sub2}Performance Metrics}
Following the notation system of \cite{SheWin:J10a}, we first write down the equivalent Fisher information matrix (EFIM) as Eq. (\ref{EFIM}).
\begin{align}\label{EFIM}
{\Je}=\sum_{n=1}^N{\lambdan\, {\RV u}_n {\RV u}_n^\mathrm{T}}
\end{align}
and the squared position error bound (SPEB), another commonly-used performance metrics for localization, is given by
\begin{align}
{\SPEB}={\tr}\Big\{{\Je}^{-1}\Big\}
\end{align}
where ${\RV u}_n=[\,\cos\thetans~~\sin\thetans]^\mathrm{T}$ and $\lambdan$ depends on the ranging errors in (\ref{ranging_exp}) as
\begin{align}\label{lambda_exp}
\lambdan=\frac{1}{\sigdis^2+\sigsyn^2}.
\end{align}

In high-accuracy localization systems, the clock-drift noise $\sigsyn^2$ can be much larger than the distance-dependent noise $\sigdis^2$ and thereby dominates (\ref{lambda_exp}).\footnote{For example, the clock drift for P410 ranging radio by Time Domain could be around 10 ns, which is much larger than the estimation error of the propagation delay with 5 GHz system bandwidth under the LOS condition (0.2 ns \cite{ChoWatWin:C07}).
Moreover, based on the experimental results \cite{LanZatPis:11}, the ranging error is hardly affected by the distance when the SNR is larger than 10dB in both indoor and outdoor environments.}
In these cases, we approximate $\lambdan$'s by a constant $\lambdaz$ for the analytical development. Meanwhile, simulation results will be provided in Section \ref{Sec9} to validate the analytical results by comparison with the case in which $\sigsyn^2$ and $\sigdis^2$ are on the same order.

Since the agent's position is unknown (so is $\thetaijs$), we formally define the LOP as follows.
\begin{definition}
Under a specific wireless network, the {\it{worst-case SPEB}} is the maximum SPEB for the agent within $\UR$, i.e., $\max_{\pos\in\UR} {\SPEB}$.
\end{definition}
\begin{definition}
{\it{Localization outage probability (LOP}}) associated with threshold $\varepsilon_\text{th}$, denoted as $\Outage(\varepsilon_\text{th})$,
is the probability that the worst-case SPEB is greater than a threshold $\varepsilon_\text{th}$ under {\it randomly} deployed networks. i.e.,
\begin{align}\label{Outageexpression}
\Outage(\varepsilon_\text{th})=\mathbb{P}
\Big\{\max\limits_{\pos\in\UR}
{\SPEB}>\varepsilon_\text{th}
\Big\}.
\end{align}
\end{definition}


\subsection{\label{Sec2Sub3}All-anchor Localization}
{We start with the case that all anchors within the agent's communication range participate in the localization. In this case, we have the following theorem.}

\begin{theorem}\label{allanchor_th}
{When all $N$ anchors in $\CR$ participate in localization, the LOP is given by}
\begin{align}\label{allanchorout}
\Outage(\varepsilon_\thr)=1-
U{\cdot}
\int_0^{\infty}{
J_1(U{\cdot}r)
\cdot J_0(r)^N \df r}
\end{align}
{where $U=\sqrt{N^2-4N/(\lambdaz\varepsilon_{\thr})}$, and $J_0(\cdot)$ and $J_1(\cdot)$ are Bessel functions of zeroth  and first order, respectively.}
\end{theorem}

\begin{IEEEproof}
Assume that the system still operate under the circumstance that $\sigsyn^2$ dominates (\ref{lambda_exp}), then (\ref{EFIM}) turns to be:
\begin{align}\label{EFIMall}
{\Je}\propto\sum_{n=1}^N{\, {\RV u}_n {\RV u}_n^\mathrm{T}}
\end{align}
Here, the constant coefficient has been reduced for simplicity. Then the localization error can be estimated as:
\begin{align}\label{eq20}
\nonumber
&\tr(\Je^{-1})\\
\nonumber
=&
\frac{
\frac{\SPEBbest}{2}
\left(
\sum\limits_{n=1}^{N}{\cos^2\thetans}+\sum\limits_{n=1}^{N}{\sin^2\thetans}
\right)
}
{\left(\sum\limits_{n=1}^{N}{\cos^2\thetans}\right)\cdot\left(\sum\limits_{n=1}^{N}{\sin^2\thetans}\right)-\left(\sum\limits_{n=1}^{N}{\cos\thetans\sin\thetans}\right)^2}\\
=&
\frac{2N}{N^2-\K^2}\SPEBbest
\end{align}
where $\K=\sqrt{
\left(\sum\limits_{n=1}^{N}{\cos 2\thetans}\right)^2+
\left(\sum\limits_{n=1}^{N}{\sin 2\thetans}\right)^2
}$.
The expression for $\K$ can also be interpreted as the distance of a two-dimensional random walk. An equivalent construction of $\K$ is provided as follows.

Suppose that there is a point staying at the origin at time slot 0. At each time slot $n$, the point has the displacement $\X_n$ which is a two-dimensional random variable. $\left|\X_n\right|$ is fixed at $1$, while its argument has the uniform distribution $U(0,2\pi)$. All $\X_n$'s are i.i.d. After $N$ time slots, the total displacement of the point is $\X=\sum\limits_{n=1}^{N}\X_n$, and the distribution of $\K=\left|\X\right|$ is the same as $\K$ in (\ref{eq20}).

Suppose the characteristic function for $\X_n$ is $\phi_{\X_n}(\omg)$, then,
\begin{align}\label{eq21}
\nonumber
&\phi_{\X_n}(\omg)=\mathbb{E}[e^{\mathrm{i}\omg\cdot\X_n}]\\
\nonumber
=&\frac{1}{2\pi}\int_0^{2\pi}e^{\mathrm{i}(\omega_1\cos\theta+\omega_2\sin\theta)}\text{d}\theta\\
\nonumber
=&\frac{1}{2\pi}\int_0^{2\pi}e^{\mathrm{i}|\omg|\cos(\theta+c_0)}\text{d}\theta\\
=&J_0(|\omg|)
\end{align}
In the equation, $J_0(\cdot)$ is the Bessel function ($0$-order). Since the addition of independent random variables can be represented as the production of their characteristic functions, we have:
\begin{align}\label{eq22}
\phi_{\X}(\omg)=J_0(|\omg|)^N
\end{align}
The density probability function for $\X$ can be obtained by applying Fourier transformation to $\phi_{\X}$.
\begin{align}\label{eq23}
\nonumber
&f_{\X}(\bm{x})\\
\nonumber
=&\left(\frac{1}{2\pi}\right)^2
\int_{\mathbb{R}^2}e^{-\mathrm{i}\cdot\omg\cdot\bm{x}}J_0(|\omg|)^N\Leb(\text{d}\omg)\\
\nonumber
=&\left(\frac{1}{2\pi}\right)^2
\int_0^{2\pi}\int_0^{\infty}e^{-\mathrm{i}r\cdot(\cos\theta\cdot{x_1}+\sin\theta\cdot{x_2})}J_0(r)^Nr\text{d}r\text{d}\theta\\
\nonumber
=&\left(\frac{1}{2\pi}\right)^2
\int_0^{\infty}{
\left[
\int_0^{2\pi}e^{-\mathrm{i}r\cdot(\cos\theta\cdot{x_1}+\sin\theta\cdot{x_2})}\text{d}{\theta}
\right]
J_0(r)^Nr\text{d}r}\\
=&\frac{1}{2\pi}
\int_0^{\infty}{
J_0(r\cdot\left|\bm{x}\right|)\cdot
J_0(r)^Nr\text{d}r}
\end{align}

The outage probability for all-anchor localization is the probability that $\tr({\Je}^{-1})$ is larger than some threshold $\varepsilon_{\text{th}}^2$. That is
\begin{align}\label{eq24}
\nonumber
&\mathbb{P}\left\{
\tr({\Je}^{-1})>\varepsilon_{\text{th}}^2
\right\}\\
\nonumber
=&\mathbb{P}\left\{
\frac{2N}{N^2-\K^2}\SPEBbest
>\varepsilon_{\text{th}}^2
\right\}\\
\nonumber
=&\mathbb{P}\left\{
\K>\sqrt{N^2-\frac{2N\SPEBbest}{\varepsilon_{\text{th}}^2}}
\right\}\\
=&1-\int_0^{U}f_{\K}(k)\text{d}k
\end{align}
where $U$ stands for $\sqrt{N^2-\frac{2N\SPEBbest}{\varepsilon_{th}^2}}$. Since $\K=\left|\X\right|$,
\begin{align}\label{eq25}
\nonumber
&\int_0^{U}f_{\K}(k)\text{d}k\\
\nonumber
=&\int_0^{U}2{\pi}k{\cdot}
\frac{1}{2\pi}
\int_0^{\infty}{
J_0(r{\cdot}k)\cdot
J_0(r)^Nr\text{d}r}
\text{d}k\\
\nonumber
=&\int_0^{\infty}{
\left[
\int_0^{U}
k{\cdot}J_0(r{\cdot}k)\text{d}k
\right]
{\cdot}J_0(r)^Nr\text{d}r}\\
=&
U{\cdot}
\int_0^{\infty}{
J_1(U{\cdot}r)
{\cdot}J_0(r)^N \text{d}r}
\end{align}
As the result, the outage probability for all-anchor localization is
\begin{align}\label{eq26}
\Outage(\varepsilon_{th}^2)=1-
U{\cdot}
\int_0^{\infty}{
J_1(U{\cdot}r)
{\cdot}J_0(r)^N \text{d}r}
\end{align}
\end{IEEEproof}

\subsection{\label{Sec2Sub4}Two-anchor Localization}
For the two-anchor localization case, the agent selects two anchors ${\anchor}_{i}$ and ${\anchor}_{j}$ for inter-node ranging measurements. Since it will yield two possible positions, we consider the one closer to $\bpos$ as the estimated position $\hpos$. Then, the corresponding SPEB of the agent is
\begin{align}\label{SPEBexpression}
{\SPEB}
={\SPEBbest}\cdot {\sin^{-2} \thetaijs}
\end{align}
where $\SPEBbest=2/\lambdaz$.
The optimal anchor selection algorithm in terms of LOP is to choose anchors ${\anchor}_{i_0}$ and ${\anchor}_{j_0}$ satisfying
\begin{align}\label{algrth1}
(i_0,j_0)=\argmin\limits_{i,j}\max\limits_{\pos\in\UR}\SPEB.
\end{align}

\begin{figure}
\centering
\includegraphics[width=0.9\columnwidth]{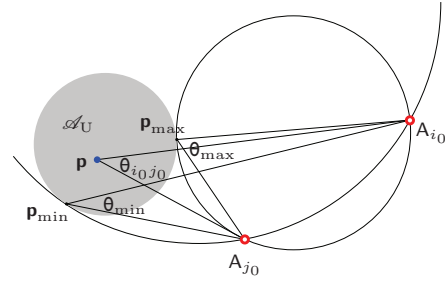}
\caption{For given $i_0$ and $j_0$, the corresponding $\pmax$ and $\pmin$ can be obtained by the tangent circles, and one of them corresponds to the worst case of $\pos$. Our anchor selection algorithm only needs to check those two positions to obtain the minimum SPEB.}
\label{algrth_fig1}
\end{figure}

Note that for each anchor pair $(\anchor_{i_0}, \anchor_{j_0})$, if the line $\anchor_{i_0}\anchor_{j_0}$ intersects $\UR$, then $\SPEB$ goes to infinity when the agent happens to be collinear with $\anchor_{i_0}\anchor_{j_0}$, which leads to an immediate outage.
Otherwise, we can draw two circles that go through both $\anchor_{i_0}$ and $\anchor_{j_0}$ and are tangent to $\UR$ at $\pmin$ and $\pmax$ (see Fig. \ref{algrth_fig1}).
By comparing the local maximum values of $\SPEB$, we can simplify the optimal selection algorithm (\ref{algrth1}) into
\begin{align}\label{algrth2}
(i_0,j_0)=\argmin\limits_{i,j} \;\max\thinspace\{\mathcal{P}(\pmax), \mathcal{P}(\pmin)\}.
\end{align}

Note that as flip ambiguity occurs only if $\anchor_{i_0}\anchor_{j_0}$ intersects $\UR$, classifying such situation as a localization outage sufficiently avoid flip ambiguity in all non-outage situations.

\section{\label{Sec3}Localization Outage Probability}

This section will derive the lower and upper bounds for the LOP given in (\ref{Outageexpression}). These bounds can provide more insights into the behavior of LOP.

\subsection{\label{Sec3Sub1}Lower Bound}
We consider the probability that all the included angles $\thetaijs$ is away from $\pi/2$ by at least $\delta$, defined as
\begin{align}\label{Pdelta}
P(\delta):=
\mathbb{P}
\left\{
\thetaijs\not\in
\Ith,
\forall i \neq j \in \Nset
\right\}
\end{align}
where $\Ith:=\left[\pi/2-\delta,\pi/2+\delta\right]$. The next theorem gives the expression or bounds for $P(\delta)$.
\begin{theorem}\label{Pdelta_th}
For $\delta \geq \pi/6$, we have
\begin{align}\label{Pdelta_case1}
P(\delta)
=N\cdot\Big(\frac{\pi-2\delta}{2\pi}\Big)^{N-1};
\end{align}
and for $\delta < \pi/6$, we have (\ref{Pdelta_case2}), shown at the top of the next page.
\begin{figure*}
\begin{align}\label{Pdelta_case2}
\Big(\frac{\pi-4\delta}{\pi}\Big)^{N-1}+(N-1)\cdot\frac{4\delta (\pi-2\delta)^{N-2}}{(2\pi)^{N-1}}
\geqslant P(\delta) \geqslant
N\cdot\Big(\frac{\pi-2\delta}{2\pi}\Big)^{N-1}+(N-2)\cdot\Big(\frac{\pi-6\delta}{2\pi}\Big)^{N-1}
\end{align}
\begin{spacing}{0.0}
\hrulefill
\end{spacing}
\end{figure*}
\end{theorem}
\begin{IEEEproof}
See Appendix A. We also attached the exact expression for $P(\delta)$ in Appendix B.
\end{IEEEproof}

Let ${\varepsilon_\thr} := \cos^{-2}(\delta) \cdot \SPEBbest$ be the threshold for the SPEB, and we next give the lower bound for the LOP.
\begin{theorem}[Lower Bound]\label{LB_th}
The LOP $\Outage\left(\varepsilon_\thr\right)$ is bounded below by $P(\delta)$.
\end{theorem}
\begin{IEEEproof}
According to (\ref{Pdelta}),
suppose that
$\thetaijs\not\in
\Ith$, $\forall i \neq j \in \Nset$,
then it is impossible for the agent to locate itself with the SPEB smaller than the threshold $\varepsilon_\thr$ by choosing only two anchors. Thus, $P(\delta)$ provides a lower bound for LOP.
\end{IEEEproof}

\subsection{\label{Sec3Sub2}Upper Bound}

First, we introduce a sub-optimal selection algorithm to relax LOP. In the new algorithm, the unknown $\thetaijs$ is approximated by $\thetaij$, i.e.,
\begin{align}\label{algrth3}
(i_0,j_0)=\argmin\limits_{i,j}\thinspace{\sin^{-2}\thetaij} \,.
\end{align}
Such algorithm will lead to near-optimal performance as $\CR$ is much larger than $\UR$, but an increased LOP. We will use (\ref{algrth3}) for deriving the upper bound for LOP.

Since the angles from anchors to $\bpos$ are no long isotropic, we introduce the concept of auxiliary communication region $\PCR := \{{\bf x}|{\bf x}\in\mathbb{R}^2, \Vert {\bf x}-{\bpos} \Vert \leqslant R+r\}$ (see Fig. \ref{figc1}), which is the union of all the possible communication regions as $\pos$ varies within $\UR$.
Suppose that the communication region is extended from $\CR$ to $\PCR$, and the corresponding outage becomes $\POutage$, which is associated with $\Outage$ according to the following proposition.

\begin{proposition}\label{PCR_lem}
For a given threshold $\varepsilon_\thr$, $\Outage$ and $\POutage$ satisfy the following relationship
\begin{align}\label{Patch1}
\Outage({\varepsilon_\thr}) \leqslant \Big(\frac{R+r}{R}\Big)^{2N} \cdot \POutage({\varepsilon_\thr})\,.
\end{align}
\end{proposition}
\begin{IEEEproof}
The probability $\POutage$ must be greater than the probability that all anchors happen to be within $\CR$ and the outage occurs, which directly leads to (\ref{Patch1}).
\end{IEEEproof}

Based on the properties of the SPEB given in (\ref{SPEBexpression}), we define $\worst_{x\in\mathscr{X}}\{\theta_x\}$ as the result $\theta_{x_0}$ such that
\begin{align}
\nonumber
x_0\in\mathscr{X}\text{,~and}~{\forall}x'{\neq}x_0,~{\left|\theta_{x'}-{\pi}/{2}\right|\leqslant\left|\theta_{x_0}-{\pi}/{2}\right|}.
\end{align}
Then, the probability $\POutage(\varepsilon_\thr)$ can be estimated as
\begin{align}\label{outageeq1}
\nonumber
&\POutage({\varepsilon_\thr})\\
\nonumber
&\quad\leqslant\mathbb{P}
\Big\{
\min\limits_{\pos\in\UR}\sin^2 \thetaizjzs \leqslant \cos^2\left(\delta\right)
\Big\}\\
&\quad=1-\int_{\Ith}
\mathbb{P}
\Big\{
\worst\limits_{\pos\in\UR}\thinspace\thetaizjzs\in
\Ith
\big|\thetaizjz=\theta
\Big\}
f(\theta)\df\theta
\end{align}
where $f(\theta)$ denotes the PDF of $\thetaizjz$ while all anchors are assumed to be distributed in $\PCR$ instead of $\CR$. Eq. (\ref{outageeq1}) divides the calculation of
$\POutage({\varepsilon_\thr})$
into two steps, which will be analyzed separately in the following.

When all anchors are located in $\PCR$, the joint distribution of $\thetaij$ is the same as that of $\thetaijs$ while all anchors are within $\CR$. Therefore, (\ref{Pdelta}) suggests
\begin{align}\label{Pdelta_PCR}
\int_{0}^{\frac{\pi}{2}-\delta}f(\theta) \df \theta+
\int_{\frac{\pi}{2}+\delta}^{\pi}f(\theta) \df \theta
=P(\delta).
\end{align}

We then define
\begin{align}\label{Q_expression}
\nonumber
\!Q_{\delta}(\theta)
&:=1-\frac{1}{(\mu-1)^2}
-\frac{1}{\mu^2}-\frac{1}{\mu^3}+\frac{2}{\mu^4}-\frac{1}{\mu^4(\mu-1)}\\
&\qquad+\frac{1}{\mu^4(\mu-1)^2}-\frac{12\log(\mu-1)}{\mu^4}
\end{align}
where $\mu=R/r\cdot\sin(\Delta\theta_{\thr})$ in which $\Delta\theta_{\thr}=\min\{|\pi/2-\delta-\theta|,|\pi/2+\delta-\theta|\}$.
The difference between $\thetaizjzs$ and $\thetaizjz$ cannot exceed the maximum value of the sum of the angular error for both edges of the angle, so that
\begin{align}\label{Angleeq1}
\nonumber
\!\!&\mathbb{P}
\Big\{
\worst\limits_{\pos\in\UR}\thetaizjzs\in
\Ith
\Big|\thetaizjz=\theta
\Big\}\\
\nonumber
&\quad\geqslant \mathbb{P}
\left\{
\left|\thetaizjzs-\theta\right|\leqslant\Delta\theta_{\thr}
\right\}\\
&\quad \geqslant \mathbb{P}
\Big\{
\frac{r}{\rv{r}_{i_0}} +
\frac{r}{\rv{r}_{j_0}}
\leqslant\sin\left(\Delta\theta_{\thr}\right)
\Big\}  = Q_{\delta}(\theta) \,.
\end{align}

\begin{theorem}[Upper Bound]\label{UB_th}
The LOP is bounded from above as
	\begin{align}
	\nonumber
		\Outage (\varepsilon_\thr ) \leqslant
		&\Big(\frac{R+r}{R}\Big)^{2N}\\
        \nonumber
		&\cdot
		\Big[
		1-Q_{\delta}(\frac{\pi}{2})-\int_{\frac{\pi}{2}}^{\frac{\pi}{2}+\delta}
		Q_{\delta}^{'}(\theta)\cdot
		P\Big(\theta-\frac{\pi}{2}\Big)
		\df \theta
		\Big]
	\end{align}
where $P(\cdot)$ and $Q_{\delta}(\cdot)$ are provided in (\ref{Pdelta_case1}), (\ref{Pdelta_case2}) and (\ref{Q_expression}).
\end{theorem}

\begin{IEEEproof}
Based on the symmetry of $f(\theta)$ and partial integration,
\begin{align}\label{outage_final}
\POutage({\varepsilon_\thr})
& \leqslant 1-2\int_{\frac{\pi}{2}}^{\frac{\pi}{2}+\delta}
Q_{\delta}(\theta)\cdot
f(\theta) \df \theta\\
\nonumber
&=1-Q_{\delta}(\frac{\pi}{2})-\int_{\frac{\pi}{2}}^{\frac{\pi}{2}+\delta}
Q_{\delta}^{'}(\theta)\cdot
P\Big(\theta-\frac{\pi}{2}\Big)
\df \theta.
\end{align}
Plugging (\ref{outage_final}) into (\ref{Patch1}) leads to the upper bound.
\end{IEEEproof}

\section{\label{Sec9}Numerical Results}

\begin{figure}[t]
\centering
\includegraphics[width = 1.05\columnwidth]{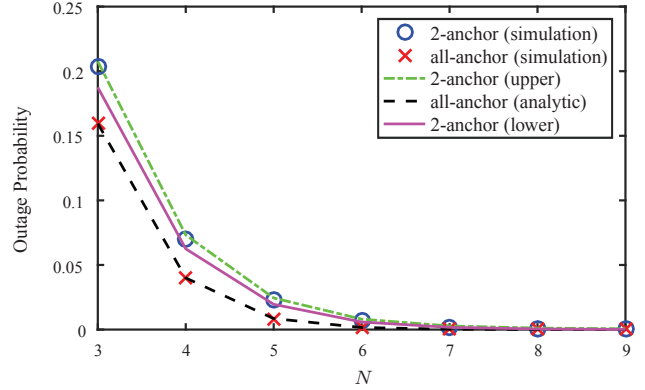}
\caption{Simulated and analytical LOP for both two-anchor and all-anchor localization, where $\varepsilon_\thr = 2\SPEBbest$ and $R/r=100$.}
\label{figc10}
\end{figure}

\begin{figure}[t]
\centering
\includegraphics[width = 1.05\columnwidth]{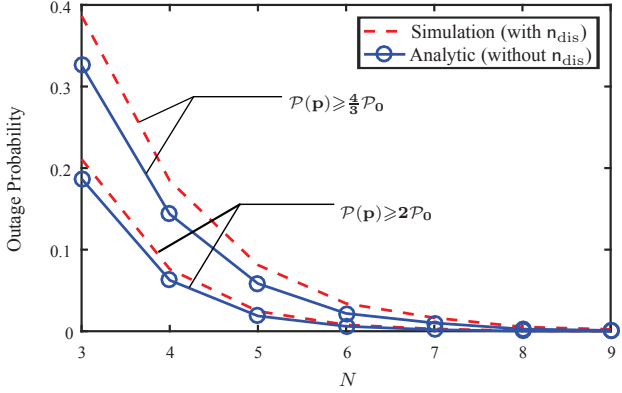}
\caption{The LOP for different SPEB threshold $\varepsilon_\thr$ and number of anchors $N$. The standard deviation of  the distance-dependent noise is $\sigdis = 0.4 \,\sigsyn$.}
\label{Pdeltafig} 
\end{figure}

This section compares the LOPs of both all-anchor and two-anchor localization through numerical evaluation and Monte Carlo simulation.

Both the simulation and analytical curves in Fig.~\ref{figc10} suggest that the LOP of the two-anchor case is close to that of the all-anchor case. However, anchor selection algorithm reduces both the energy consumption and ranging time to only $2/N$ of that in the all-anchor case. This is particularly important for low-power and delay-sensitive applications.

In the two-anchor case, when $R \gg r$, the difference between $\thetaij$ and $\thetaijs$ is negligible and thus the upper bound in (\ref{outage_final}) coincides with the lower bound $P(\delta)$, which can thereby serve as a tight approximation for the LOP. Fig.~\ref{Pdeltafig} depicts the curves of $P(\delta)$ with respect to $N$ for different SPEB thresholds, where the communication region is sufficiently large compared to $\UR$. We also consider distance-dependent noises and let $\sigdis = 0.4 \, \sigsyn$. The comparison in Fig.~\ref{Pdeltafig} shows that the analytical results can be used as a good approximation to characterize the LOP when the clock drift noise is not dominant.

\begin{figure}[t]
\centering
\includegraphics[width = 1.05\columnwidth]{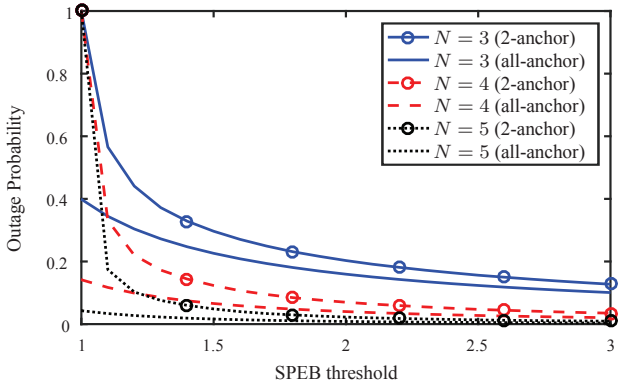}
\caption{The unit for the horizontal axis is $\SPEBbest$. The figure displays the relationship between the SPEB thresholds and LOPs for $N=3$, $4$ and $5$.}
\label{rslt}
\end{figure}

For a fixed number of anchors ($N=3,4,5$), Fig.~\ref{rslt} shows the LOPs as a function of the SPEB threshold. As it stands, the LOPs for both the two-anchor and all-anchor cases decrease with the threshold, and the decrease rate is large when the threshold is close to $\SPEBbest$. Moreover, the gap between the LOPs for the two-anchor and all-anchor case decreases with the SPEB threshold and the number of anchors, implying that anchor selection yields closer-to-optimal outage performance when the accuracy guarantee is less stringent and the network size is larger.


%

\section{\label{Sec10}Conclusion}
This paper derived the outage probability of self-localization due to the randomness of network geometry. We first determined the exact LOP when all the neighboring anchors are used for localization, and then upper and lower bounds for the LOP when the best pair of anchors are selected for localization.
Numerical results demonstrated the tightness of the bounds, and showed the advantage of the two-anchor localization in terms of efficiency. Our paper highlights the tradeoffs between outage probability and accuracy guarantee for two-anchor and all-anchor cases, which will facilitate the design of efficient localization systems via anchor selection.

\bibliographystyle{IEEEtran}

\section*{\label{App_A}Appendix A}
This section deduces the (approximate) expressions for $P(\delta)$, which reflects the outage caused by the anchor deployment in the two-anchor situation.

Let $\thetanp=2\thetans$ (in the sense of wraparound) for all $n\in\Nset$,
and $\thetaijp$ is likely defined as the included angle between $\thetaip$ and $\thetajp$. Then ($\ref{Pdelta}$) can be transformed into
\begin{align}\label{Pdelta2}
P(\delta)=\mathbb{P}
\left\{
{\forall}i,j,
\thetaijp\in
\left[
0,\pi-2\delta
\right]
\right\}
\end{align}

Following passages are divided into two subsections. The first subsection copes with the precise expression for the case of $\delta{\geqslant}\frac{\pi}{6}$, while the second subsection only calculate the upper and lower bounds for the case of $\delta<\frac{\pi}{6}$.

\subsection{\label{Appenddix_A_1}For $\delta\geq\frac{\pi}{6}$}
\begin{figure}
\centering
\psfrag{A1}{\scriptsize$\overset{\frown}{A_1}$}
\psfrag{A2}{\scriptsize$\overset{\frown}{A_2}$}
\psfrag{th'_k}{\scriptsize$\thetakp$}
\psfrag{th'_i0}{\scriptsize${\rv{\theta}_{i_0}'}$}
\psfrag{th'_j0}{\scriptsize${\rv{\theta}_{j_0}'}$}
\includegraphics[width=3.5 in]{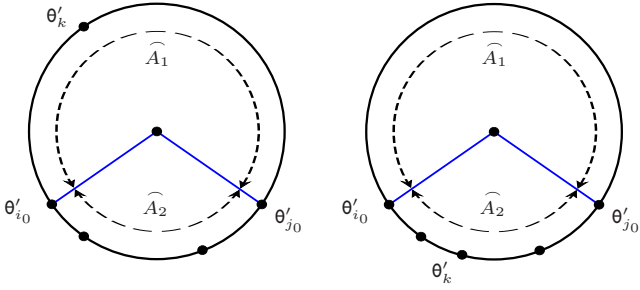}
\caption{Left figure corresponds to the situation that $\theta'_k\in\overset{\frown}{A_1}$, while the right one corresponds to $\theta'_k\in\overset{\frown}{A_2}$.}
\label{figc_new1}
\end{figure}

\begin{theorem}\label{angle_th}
If $\delta\geq\frac{\pi}{6}$ and $\forall i,j\in\Nset$,$\thetaijp < \pi-2\delta$, then $\exists i_0,j_0$, s.t. $\thetaizjzp=\max\limits_{i,j}\thetaijp$, and $\forall k\not=i_0,j_0, \thetaizjzp={\rv{\theta}_{i_0k}'}+{\rv{\theta}_{j_0k}'}$.
\end{theorem}
\begin{IEEEproof}
Let $(i_0,j_0)=\arg\max\limits_{i,j}\thetaijp$ (see Fig. \ref{figc_new1}). If $\thetakp$ lies within arc $\overset{\frown}{A_1}$, as shown in Fig. \ref{figc_new1} (left), recall that $\thetaizjzp$ is the largest included angle, then both ${\rv{\theta}_{i_0k}'}$ and ${\rv{\theta}_{j_0k}'}$ cannot include $\thetaizjzp$. Therefore, the included angles between ${\rv{\theta}_{i_0}'}$, ${\rv{\theta}_{j_0}'}$ and $\thetakp$ are exclusive to each other. Consider that $\delta\geq\frac{\pi}{6}$, each included angle must be smaller than $\pi-2\delta\leqslant\frac{2}{3}\pi$ and the sum of them, the central arc of the whole circle, is smaller than $2\pi$, which causes the contradiction. Hence all $\thetakp$ must be within arc $\overset{\frown}{A_2}$, as shown in Fig. \ref{figc_new1} (right). Theorem \ref{angle_th} is proved.
\end{IEEEproof}

\begin{figure}
\centering
\psfrag{0,2pi}{\scriptsize$0,2\pi$}
\psfrag{pi}{\scriptsize$\pi$}
\psfrag{pi/2}{\scriptsize$\frac{1}{2}\pi$}
\psfrag{pi3/2}{\scriptsize$\frac{3}{2}\pi$}
\psfrag{pi-2d}{\scriptsize$\pi-2\delta$}
\psfrag{th_min}{\scriptsize${\rv{\theta}_{\min}'}$}
\psfrag{th_1}{\scriptsize${\rv{\theta}_{1}'}$}
\includegraphics[width=2 in]{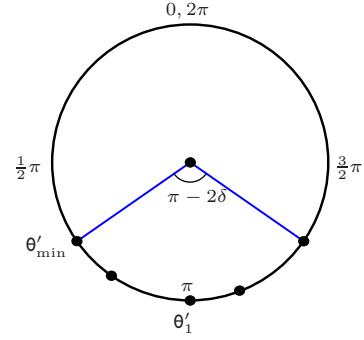}
\caption{When the smallest angle ${\rv{\theta}_{\min}'}$ is determined, all $\thetaip$'s are neither smaller than ${\rv{\theta}_{\min}'}$ nor larger than ${\rv{\theta}_{\min}'}+\pi-2\delta$.}
\label{figc_new2}
\end{figure}

To distinguish different positions on the circle, we assign the radian measure to the circle circumference counter-clockwisely. Without the loss of generality, we fix ${\rv{\theta}_1'}=\pi$ (see Fig. \ref{figc_new2}), then the smallest angle ${\rv{\theta}_{\min}'}=\min\limits_{i}\thetaip$ has the following distribution:
\begin{align}\label{new_eq2}
f_{\min}(x)\text{d}x=
\begin{cases}
    \frac{(N-1){(2\pi-x)}^{N-2}}{(2\pi)^{N-1}}\text{d}x, &2\delta<x<\pi\\
    \left(\frac{1}{2}\right)^{N-1}, &x=\pi\\
    0, &\text{Otherwise}
\end{cases}.
\end{align}
As long as ${\rv{\theta}_{\min}'}$ is determined, all $\thetaip$'s are neither smaller than ${\rv{\theta}_{\min}'}$ nor larger than ${\rv{\theta}_{\min}'}+\pi-2\delta$. Therefore, the probability that $P(\delta)$ can be precisely expressed as
\begin{align}\label{new_eq3}
\nonumber
P(\delta)
=&\int_{2\delta}^{\pi}
\frac{(N-1){(2\pi-x)}^{N-2}}{(2\pi)^{N-1}}
\left(\frac{\pi-2\delta}{2\pi-x}\right)^{N-2}
~\text{d}x\\
\nonumber
&+\left(\frac{1}{2}\right)^{N-1}\cdot
\left(\frac{\pi-2\delta}{\pi}\right)^{N-1}\\
=&N\cdot\left(\frac{\pi-2\delta}{2\pi}\right)^{N-1}.
\end{align}

\subsection{\label{Appenddix_A_2}For $\delta<\frac{\pi}{6}$}
This subsection will provide the upper and lower bound of $P(\delta)$ in simple forms. The exact expression for $P(\delta)$ in this case is quite complex and will be deduced in Appendix B.

When $2\delta\geq{\rv{\theta}_{\min}'}$, the event in $P(\delta)$ does not hold since the included angle between ${\rv{\theta}_1'}$ and ${\rv{\theta}_{\min}'}$ has already exceeded $\pi-2\delta$.

\begin{figure}
\centering
\psfrag{0,2pi}{\scriptsize$0,2\pi$}
\psfrag{pi}{\scriptsize$\pi$}
\psfrag{th_min}{\scriptsize${\rv{\theta}_{\min}'}$}
\psfrag{th_1}{\scriptsize${\rv{\theta}_{1}'}$}
\psfrag{x}{\scriptsize$x$}
\psfrag{I1}{\scriptsize$\Ia$}
\psfrag{I2}{\scriptsize$\Ib$}
\psfrag{I3}{\scriptsize$\Ic$}
\psfrag{2pi-2d}{\scriptsize$2\pi-2\delta$}
\psfrag{x+pi+2d}{\scriptsize$x+\pi+2\delta$}
\psfrag{x+pi-2d}{\scriptsize$x+\pi-2\delta$}
\includegraphics[width=2 in]{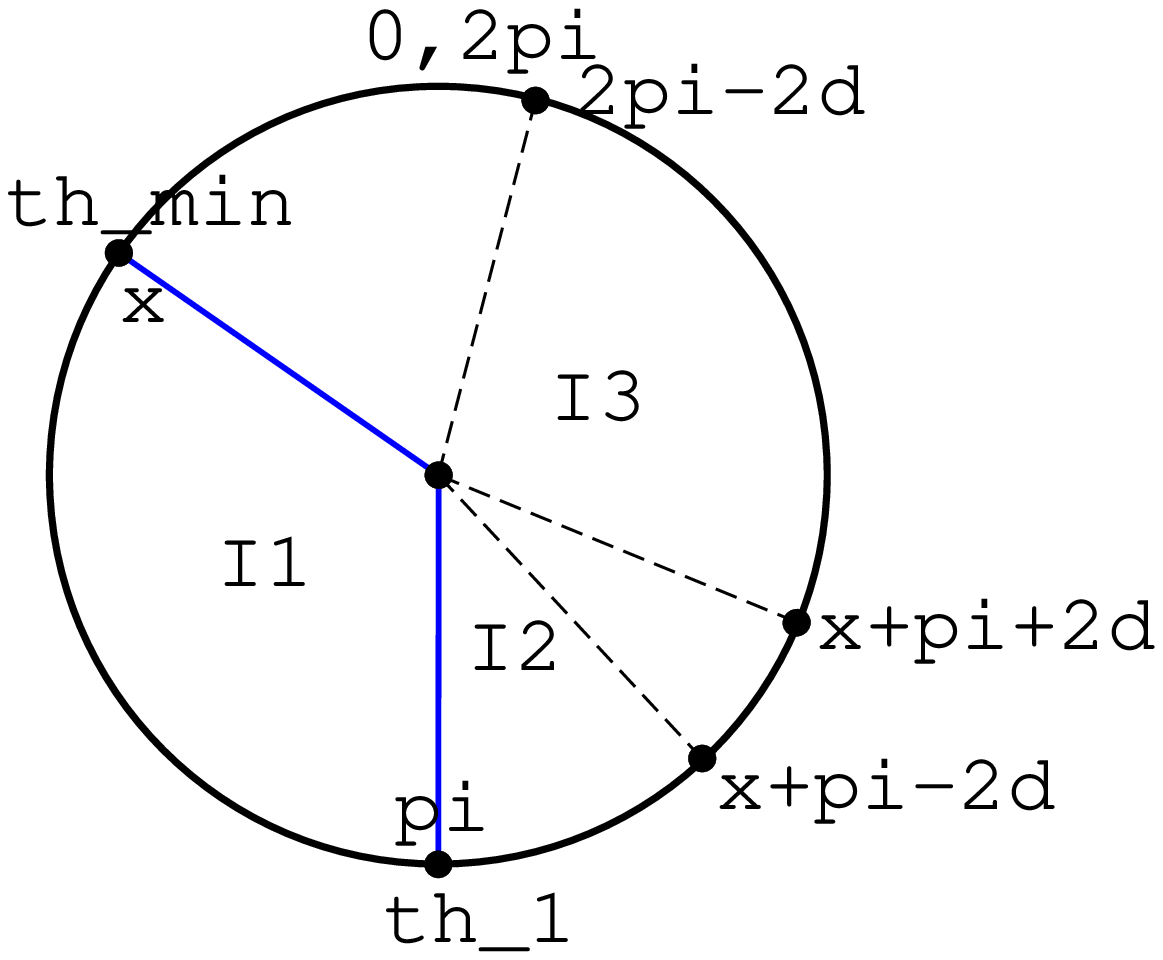}
\caption{When ${\rv{\theta}_{\min}'}$ and ${\rv{\theta}_1'}$ are fixed and $2\delta{\leqslant}{\rv{\theta}_{\min}'}<\pi-4\delta$, the necessity of the outage condition is that all other $\thetaip$'s stay within $\Ia$, $\Ib$ or $\Ic$.}
\label{figc_new3}
\end{figure}

When $2\delta<{\rv{\theta}_{\min}'}<\pi-4\delta$, the rest $N-2$ $\thetaip$'s can be placed in 3 intervals in order to satisfy the event in $P(\delta)$: $\Ia=[x,\pi)$, $\Ib=[\pi,x+\pi-2\delta]$, $\Ic=[x+\pi+2\delta,2\pi-2\delta]$, see Fig. \ref{figc_new3}. Notice that the points in $\Ib$ will never conflict with other points in $\Ia$ or $\Ic$ and the points in the same interval also do not conflict with each other. So we consider about the following two situations:
\begin{enumerate}[(1)]
\item
Either $\Ia$ or $\Ic$ is empty. The conditional probability given $\theta'_{\min}$ for this situation is denoted as $P_1$.
\item
All the rest $\theta'_i$'s can be placed in all three intervals, and the conflicts are ignored. The conditional probability given $\theta'_{\min}$ for this situation is denoted as $P_2$.
\end{enumerate}
The first situation is the sufficient (but not necessary) condition of the outage condition, while the second condition is necessary but not sufficient. Hence, $P_1$ and $P_2$ can be relied on to achieve the lower bound and upper bound of the outage probability respectively. We have:
\begin{align}\label{new_eq4}
\nonumber
&\left\{
   \begin{array}{l}
    P_1(x)=\left(\frac{\Leb(\Ia{\cup}\Ib)}{2\pi-x}\right)^{N-2}+
    \left(\frac{\Leb(\Ib{\cup}\Ic)}{2\pi-x}\right)^{N-2}-
    \left(\frac{\Leb(\Ib)}{2\pi-x}\right)^{N-2}\\
    P_2(x)=\left(\frac{\Leb(\Ia{\cup}\Ib{\cup}\Ic)}{2\pi-x}\right)^{N-2}
   \end{array}
\right.\\
&\Rightarrow
\left\{
   \begin{array}{l}
    P_1(x)=\frac{(\pi-2\delta)^{N-2}+(\pi-6\delta)^{N-2}-(x-2\delta)^{N-2}}{(2\pi-x)^{N-2}}\\
    P_2(x)=\left(\frac{2\pi-6\delta-x}{2\pi-x}\right)^{N-2}
   \end{array}
\right.
\end{align}
In equation (\ref{new_eq4}), $\Leb(\cdot)$ is the Lebesgue measure, and the expression of $P_1$ results from the inclusion–exclusion principle.

When $\pi-4\delta{\leqslant}{\rv{\theta}_{\min}'}<\pi$, all the rest $\thetaip$'s must be within $[x,x+\pi-2\delta]$ in order to avoid the conflict, so that the conditional probability turns to be
\begin{align}\label{new_eq5}
\mathbb{P}\left\{\thetaijp<\pi-2\delta,\forall i,j\in{\Nset}|{\rv{\theta}_{\min}'}=x\right\}
=\left(\frac{\pi-2\delta}{2\pi-x}\right)^{N-2}.
\end{align}

When ${\rv{\theta}_{\min}'}=\pi$, then the conditional probability is
\begin{align}\label{new_eq6}
\mathbb{P}
\left\{\thetaijp<\pi-2\delta,\forall i,j\in\Nset|{\rv{\theta}_{\min}'}=\pi\right\}
=\left(\frac{\pi-2\delta}{\pi}\right)^{N-1}.
\end{align}

With the help of (\ref{new_eq4}) to (\ref{new_eq6}), both the lower and upper bound for $P(\delta)$ can be obtained as (\ref{new_eq7}) and (\ref{new_eq8}), and thereby (\ref{Pdelta_case2}) holds.
\begin{figure*}
\begin{align}\label{new_eq7}
\nonumber
P(\delta){\geqslant}&\int_{2\delta}^{\pi-4\delta}{f_{\min}(x){\cdot}P_1(x)\text{d}x}
+\int_{\pi-4\delta}^{\pi}{f_{\min}(x){\cdot}\left(\frac{\pi-2\delta}{2\pi-x}\right)^{N-2}\text{d}x}
+\left(\frac{1}{2}\right)^{N-1}{\cdot}\left(\frac{\pi-2\delta}{\pi}\right)^{N-1}\\
=&N\cdot\left(\frac{\pi-2\delta}{2\pi}\right)^{N-1}+(N-2)\cdot\left(\frac{\pi-6\delta}{2\pi}\right)^{N-1}
\end{align}
\begin{align}\label{new_eq8}
\nonumber
P(\delta){\leqslant}&\int_{2\delta}^{\pi-4\delta}{f_{\min}(x){\cdot}P_2(x)\text{d}x}
+\int_{\pi-4\delta}^{\pi}{f_{\min}(x){\cdot}\left(\frac{\pi-2\delta}{2\pi-x}\right)^{N-2}\text{d}x}
+\left(\frac{1}{2}\right)^{N-1}{\cdot}\left(\frac{\pi-2\delta}{\pi}\right)^{N-1}\\
=&\left(\frac{2\pi-8\delta}{2\pi}\right)^{N-1}+(N-1)\cdot\frac{(\pi-2\delta)^{N-2}4\delta}{(2\pi)^{N-1}}
\end{align}
\begin{align}\label{new_eq11}
\nonumber
P(\delta)=&\int_{2\delta}^{\pi-4\delta}f_{\min}(x){\cdot}
\sum\limits_{n=0}^{\scriptscriptstyle N-2}
\sum\limits_{m=0}^{\scriptscriptstyle N-2-m}
\binom{\scriptstyle N-2}{\scriptstyle n}\binom{\scriptstyle N-2-n}{\scriptstyle m}
\Big(\frac{\pi-x-4\delta}{2\pi-x}\Big)^{\scriptscriptstyle n}
\Big(\frac{\pi-x}{2\pi-x}\Big)^{\scriptscriptstyle m}
\Big(\frac{x-2\delta}{2\pi-x}\Big)^{\scriptscriptstyle N-2-n-m}
{\mathbb{E}}\left[\left(\frac{\pi-x-\Y}{\pi-x}\right)^m\right]
\text{d}x\\
\nonumber
&+\int_{\pi-4\delta}^{\pi}{f_{\min}(x){\cdot}\left(\frac{\pi-2\delta}{2\pi-x}\right)^{N-2}\text{d}x}
+\left(\frac{1}{2}\right)^{N-1}{\cdot}\left(\frac{\pi-2\delta}{\pi}\right)^{N-1}\\
\nonumber
=&\frac{N-1}{(2\pi)^{N-1}}\cdot
\sum\limits_{n=0}^{\scriptscriptstyle N-2}
\sum\limits_{m=0}^{\scriptscriptstyle N-2-n}
\binom{\scriptstyle N-2}{\scriptstyle n}
\binom{\scriptstyle N-2-n}{\scriptstyle m}
\int_{2\delta}^{\pi-4\delta}
(\pi-x-4\delta)^{\scriptscriptstyle n}
(x-2\delta)^{\scriptscriptstyle N-2-n-m}
{\mathbb{E}}\left\{(\pi-x-\Y)^{\scriptscriptstyle m}\right\}
\text{d}x\\
&+\frac{(\pi-2\delta)^{N-2}}{(2\pi)^{N-1}}
[(N-1)\cdot{4\delta}+\pi-2\delta]
\end{align}
\begin{spacing}{0.0}
\hrulefill
\end{spacing}
\end{figure*}

\section*{\label{App_B}Appendix B}
This section discusses about the exact expression of $P(\delta)$ in the case of $\delta<\frac{\pi}{6}$. Before going further, we first think about the following interval coverage problem.

Suppose that we have the interval $\mathcal{I}=[0,L]$, $n$ random points are placed within $\mathcal{I}$. Their positions $\Xcover_1,\Xcover_2,\cdots,\Xcover_n$ are $n$ i.i.d. uniform random variables. Each point $\Xcover_i$ generates a smaller interval with length $D$ centered at $\Xcover_i$. The union of all $n$ small intervals is denoted as $\Sset=\bigcup\limits_{i=1}^{n}\{x\big||x-\Xcover_i|<\frac{D}{2}\}$, then we want to find out the exact expression of the measure of $\Sset$.

Let $\Y=\Leb(\Sset)$, then on the basis of literatures about linear coverage \cite{MomCov:Gre}\cite{ProDis:Vot}, the probability density function for $\Y$ can be expressed as:
\begin{align}\label{new_eq9}
\nonumber
&f_\Y(y)=n\sum\limits_{j=0}^{\left[\frac{y}{D}\right]-1}\sum\limits_{r=0}^{\left[\frac{y}{D}\right]-j-1}(-1)^r{\binom{n-1}{j}}{\binom{n-1}{j+1}}\cdot\\
&{\binom{n-j-1}{r}}\Big(1-\frac{y-D}{L}\Big)^{j+1}[y-D(j+r+1)]^{n-j-2}
\end{align}
When $n{\geqslant}1$, $f_\Y(y)$ only takes value while $y\in[D,\min(L+D,n{\cdot}D)]$. When $n=0$, $f_\Y(y)={\delta}(y)$. The ${\delta}(y)$ here is the Dirac-Delta function, which differs from the notation of the angle $\delta$.

Back to the original problem, as aforementioned, in the case of $\delta<\frac{\pi}{6}$, when the value of ${\rv{\theta}_{\min}'}$ is given (denoted as $x$) and ${\rv{\theta}_{\min}'}<2\delta$, all the remaining $N-2$ $\thetaip$'s must be within intervals $\Ia$, $\Ib$, and $\Ic$. Given that there are $n$ points falling within $\Ic$, $m$ points within $\Ia$ and $N-2-n-m$ points within $\Ib$, then the conditional probability of the outage condition under this setting can be expresses as
$\frac{1}{(\pi-x)^m}{\cdot}{\mathbb{E}}\left[(\pi-x-\Y)^m\right]$, where $\S$ has the same meaning as the problem raised above. Here, $L$ is the measure of $\Ic$ and $D=4\delta$. Based on the PDF of $\Y$, this conditional probability is:
\begin{align}\label{new_eq10}
{\mathbb{E}}\left[(\pi-x-\Y)^m\right]=\int_{0}^{\pi-x}(\pi-x-y)^m{\cdot}f_\Y(y)\text{d}y
\end{align}
The total outage probability can be expressed in equation (\ref{new_eq11}).

Equation (\ref{new_eq11}) is quite complex and it is unknown that if the expression can be reduced to the simpler form. In this paper, we will not use this expression to analysis the localization problem due to its complexity. Instead, inequalities (\ref{new_eq7}) and (\ref{new_eq8}) are much simpler and it will be shown in later sections that they are sufficiently closed to the simulation result. Equation (\ref{new_eq11}) just serves as the theoretic analysis which makes the discussion complete.

\ifCLASSOPTIONcaptionsoff
  \newpage
\fi

\end{document}